\documentstyle[preprint,aps]{revtex}
 
\def\beq{\begin{equation}}
\def\eeq{\end{equation}}
\def\be{\begin{eqnarray}}
\def\ee{\end{eqnarray}}
\def\ci{\cite}
\def\bi{\bibitem}

\def\magk{|{\bf k}|}

\begin{document}


\draft

\title{Backward hadron production in neutrino-nucleus
interactions}

\author
{ O. Benhar$^{1}$\cite{byline},  S. Fantoni${^2}$, G.I. Lykasov${^3}$ }

\address
{ $^1$ Jefferson Laboratory, Newport News, VA 23606, USA \\
and \\ Department of Physics, Old Dominion University,
Norfolk, VA 23529, USA \\
$^2$ International School for Advanced Studies (SISSA)\\
Via Beirut, 2-4. I-30014 Trieste, Italy\\
$^3$ Joint Institute for Nuclear Research, Dubna 141980, Moscow
Region, Russia }

\date{\today}

\maketitle

\begin{abstract}

The production of backward pions in lepton-nucleus
collisions is analyzed. We show that
a large yield of high momentum backward pions
can be explained by the Regge asymptotic of
the distribution of nucleons carrying a large momentum fraction
in the nuclear target. The calculated
spectra of pions emitted in the $\nu + Ne \rightarrow \mu^- + \pi^{\pm} 
+ X$ reaction are in satisfactory agreement
with the available experimental data.

\end{abstract}
 

Semi-inclusive lepton-induced hadron production reactions in which
the detected hadron is emitted backward in the rest frame of the target
have long been recognized as a powerful tool to investigate short-range 
nuclear dynamics. The studies recently carried out in refs. \ci{bfl1,bfl4}, 
focused 
on the emission of backward protons in $e$-A scattering, have shown that: 
i) the inclusion of high momentum components in the nucleon momentum 
distribution,  
$n(\magk)$, is needed to account for the available data and ii) the behavior 
of $n(\magk)$ at large $\magk$ can be obtained from the 
Regge asymptotic of the proton spectra observed in soft hadron-nucleus
collisions, provided the nonpertubative $Q^2$-dependence of the
quark distribution in the nucleus is taken into account \ci{bfl4}.

Besides providing information on the structure of 
the nuclear wave function at short interparticle distance, the study 
of backward pion production in semi-inclusive lepton-nucleus
processes allows to quantitatively investigate the 
fragmentation of the target nucleus into hadrons. In this paper we
describe a theoretical calculation of the spectra
of backward $\pi^{\pm}$ arising from target fragmentation 
in neutrino-nucleus interactions, and compare the results to 
the experimental data of ref.\ci{wa59}. 

The semi-inclusive spectrum of hadrons (e.g. pions) of energy $E_h$
and three-momentum ${\bf p}$, detected in coincidence with a lepton
of energy $E^\prime$ emitted
in the direction specified by the solid angle $\Omega$, can be written, 
within the framework of the impulse approximation, as
\be
\nonumber
\rho_{\ell A \rightarrow \ell^\prime h X}(q,{\bf p}) & = &
E_h \frac{d\sigma} {d\Omega dE^\prime d{\bf p}} \\
 & = & \int d^4k S(k) r(k) \left[
\frac{Z}{A} \rho_{\ell p \rightarrow \ell^\prime h X}(k,q,{\bf p}) +
\frac{N}{A} \rho_{\ell n \rightarrow \ell^\prime h X}(k,q,{\bf p})
\right]\ .
\label{1}
\ee
In the above equation $q$ is the four-momentum transferred
by the lepton, $S(k)$ is 
the relativistic-invariant function describing the nuclear vertex
with an outgoing virtual nucleon of four-momentum $k$,
$\rho_{\ell p \rightarrow \ell^\prime h X}$ and
$\rho_{\ell n \rightarrow \ell^\prime h X}$ are
the semi-inclusive spectra of hadrons
 $h$ produced in $\ell p$ and $\ell n$
collisions, respectively, $Z$ is the charge of the nucleus, $A$ its
atomic number and $N=A-Z$ the number of neutrons.
The quantity $r(k)$ is the ratio of the
fluxes associated with $\ell N$ ($N=p,n$) and $\ell A$ collisions.

The spectrum of eq.(\ref{1}) can be rewritten in terms of the
relativistic-invariant variable
\beq
z = \frac{M_A}{m}\frac{(pq)}{(P_A q)}\ ,
\label{3}
\eeq
where $p \equiv (E_h,{\bf p})$ is the four-momentum of the emitted hadron, 
$m$ is the nucleon mass and $M_A$ and $P_A$ denote the target mass
and four-momentum, respectively. The resulting expression is 
\be
\nonumber
\rho_{\ell A \rightarrow \ell^\prime h X}(x,Q^2;z,p_t)
 & = & \int_{z\leq y} dy\ d^2k_t\  f_A(y,Q^2,k_t) \left[
\frac{Z}{A}\rho_{\ell p \rightarrow \ell^\prime h X}
\left(\frac{x}{y},Q^2;\frac{z}{y},p_t-k_t\right) \right. \\
& + & \left.
\frac{N}{A}\rho_{\ell n \rightarrow \ell^\prime h X}
\left(\frac{x}{y},Q^2;\frac{z}{y},p_t-k_t\right)\right]\ .
\label{4}
\ee
In the above equation, $Q^2 = -q^2$, $x$ is the Bjorken scaling variable,
$p_t=|{\bf p}|^2 - p_z^2$ and $k^2_t = |{\bf k}|^2 - k_z^2$, $p_z$
and $k_z$ being the components of ${\bf p}$ and ${\bf k}$ along the
direction of ${\bf q}$. The distribution function $f_A(y,Q^2,k_t)$ is
defined as
\beq
f_A(y,Q^2,k_t) = \int dk_0\ dk_z\ S(k)\ y\
\delta \left( y- \frac{M_A}{m} \frac{(k q)}{(P_A q)} \right)\ .
\label{5}
\eeq
Note that the above definition includes the dependence of $f_A(y,Q^2,k_t)$
upon $Q^2$, which was neglected in refs.\ci{bfl1,bfl4} assuming the
validity of the Bjorken limit. Hence, eq.(\ref{5}) can be safely
used at any value of $Q^2$, including the region
$Q^2 < 50$ (GeV/c)$^2$, where the $Q^2$-dependence of $f_A(y,Q^2,k_t)$
has been shown to be sizeable \cite{bfl2,bfl3}.

When the produced hadron is emitted backward, eq.(\ref{4})  
can be rewritten in the simplified form
\be
\nonumber
\rho_{\ell A \rightarrow \ell^\prime h X}(x,Q^2;z) & = &
\int_{z\leq y} dy f_A(y,Q^2) \left[
\frac{Z}{A}\rho_{\ell p \rightarrow \ell^\prime h X}
\left(\frac{x}{y},Q^2;\frac{z}{y} \right) \right. \\
& + & \left.
\frac{N}{A}\rho_{\ell n \rightarrow \ell^\prime h X}
\left(\frac{x}{y},Q^2;\frac{z}{y} \right) \right]\ ,
\label{6}
\ee
with
\beq
f_A(y,Q^2) = \int d^4k\ S(k)\ y\
\delta\left( y - \frac{M_A}{m}\frac{(k q)} {(P_A q)} \right)\ .
\label{7}
\eeq

In principle, the distribution function $f_A(y,Q^2)$ can
be calculated within nuclear many-body theory, approximating 
$S(k)$ with the nonrelativistic spectral function
$P(k)$, yielding the probability of finding a nucleon with momentum
${\bf k}$ and removal energy $(m - k_0)$ in the target nucleus \ci{fs}.
However, due to the limited range of momentum and removal energy covered by
nonrelativistic calculations of $P(k)$ 
(typically $\magk <$ .7 GeV/c and $(m - k_0) <$ .6 GeV.
 See e.g. ref.\ci{bf1}), 
this procedure can only be used in the region $y < 1.7$.
An alternative approach to obtain $f_A(y,Q^2)$ at larger $y$, 
based on the calculation of the overlap of the relativistic-invariant 
phase-space available to quarks belonging to 
strongly correlated nucleon, has been recently proposed in ref.\ci{bfl1}. 
Within this approach 
the asymptotic behaviour of $f_A(y,Q^2)$ at 
$y > 1.7$ and small $Q^2$ can be related to the
Regge asymptotic of the constituent quarks distribution at  
$y \rightarrow 1$ and the nonpertubative
$Q^2$-dependence is taken into account following 
ref.\ci{kaid3}.  
  
The second ingredient entering the calculation
of the spectrum defined by eq.(\ref{4}), i.e. the elementary
semi-inclusive spectrum  
$\rho_{\ell N \rightarrow \ell^\prime h X}(k,q,{\bf p})$, 
can be evaluated using the approach developed in refs.
\ci{kaid1,kaid2}. According to 
\ci{kaid1}, the elementary production process can be described 
in terms of planar and cylindrical graphs in the $s$-channel, 
as shown in fig. 1 for the case of neutrino interactions, in which
the exchanged particle is a $W$-boson.
The planar graph of fig.1a describes the 
scattering of the incoming neutrino off a valence quark.
Therefore, the corresponding contribution to the spectrum of fast 
backward hadrons, that will be denoted $F_P^N(x,Q^2;z)$, 
is proportional to the valence quark distribution multiplied by 
the fragmentation function of the spectator diquark into the detected
hadron. In ref.\ci{diper} $F_P^N(x,Q^2;z)$ has been evaluated 
for the case of proton production in neutrino scattering, whereas 
the case of pion production in electron scattering has been 
discussed in ref.\ci{bfl1}. 
The planar graph contribution to the process 
$\nu + N \rightarrow \mu^- + \pi + X$ reads
\be
F_P^N(x,Q^2;z) = z\ \phi_1(x,Q^2)
\left[ \frac{1}{3} D_{uu \rightarrow \pi} \left( \frac{z}{1-x} \right) +
\frac{2}{3} D_{ud \rightarrow \pi} \left( \frac{z}{1-x} \right) 
\right]\ ,
\label{15}
\ee
where
\be
\phi_1(x,Q^2) = \frac{G^2 m E}{\pi} \frac{x}{1-x}
\left( \frac{m^2_W}{m^2_W+Q^2} \right) d_v(x,Q^2)\ .
\label{16}
\ee
In the above equation
$G$ is the Fermi coupling constant, $E$ is the energy of 
the incoming neutrino, $x$ is the Bjorken variable and
$d_v$ is the $d$-quark distribution.
The functions $D_{uu \rightarrow \pi}$ and $D_{ud \rightarrow \pi}$ 
describe the fragmentation
of the spectator diquark, $uu$ or $ud$, into a positive or negative pion,
whereas $m_W$ is the $W$-boson mass. 

It is well known that, in addition to the dependence upon $x$ and
$z/(1-x)$, the quark distributions and fragmentation
functions display a $Q^2$ dependence. For small $Q^2$
they exhibit true Regge asymptotic \ci{kaid1,kaid2} at 
$x_F\rightarrow \pm 1$, $x_F=2 p_{z}/W_X$ and $W_X$ being the Feynman 
variable and the invariant mass of the undetected debris, respectively. 
This Regge asymptotic, discussed in the Appendix, is mainly determined 
by the intercepts of the Reggeon ($\alpha_R(0)$) and
the averaged baryon 
trajectories (${\widetilde \alpha}_B(0))$ and is completely different from 
the corresponding behaviour observed in deep inelastic
lepton-nucleon scattering, where  $Q^2$ is large \ci{kaid1,kaid2}. 
According to refs.\ci{kaid1} and \ci{kaid2} the planar graph of fig.1a 
corresponds 
to the one-Reggeon graph in the $t$-channel, having the asymptotic behaviour 
$W^{\alpha_R(0)-1}_X$, where $\alpha_R(0)=1/2$ is the Reggeon
intercept. As a consequence, the contribution of this graph is a decreasing 
function of $W_X$.
  
The contribution of the so called cylindrical graph, associated with
scattering off a sea quark and shown in 
fig.1b, will be denoted $F_C(x,Q^2;z)$. It can be written 
in the form \ci{bfl1,kaid2,ls}
\be
F_C(x,Q^2;z) = z\ \phi_2(Q^2) \left[ L_1(z,x) + L_2(z,x) \right]\ ,
\label{17}
\ee 
where
\be
\phi_2(Q^2)=\frac{G^2 m E}{\pi} \left( \frac{m^2_W}{m^2_W+Q^2} \right)\ ,
\ee
\be
L_1 = \int_z^{1-x} \left[ u_v(y)D_{u_v \rightarrow \pi}
\left( \frac{z}{y} \right) +
d_v(y)D_{d_v \rightarrow \pi}\left( \frac{z}{y} \right) \right]\ ,
\frac{dy}{y}
\label{18}
\ee
and
\be
L_2 = \int_z^{1-x} \left\{ \frac{4}{3} f_{ud}(y) D_{ud \rightarrow \pi}
\left( \frac{z}{y} \right) +
\frac{1}{3}\left[ f_{uu}D_{uu \rightarrow \pi}
\left( \frac{z}{y} \right)+f_{dd}(y)
D_{dd \rightarrow \pi} \left( \frac{z}{y} \right) \right] 
\right\} \frac{dy}{y}\ .
\label{19}
\ee
In the above equations $u_v$ is the 
distribution of the valence $u$-quark, whereas $f_{uu},f_{ud}$ and 
$f_{dd}$ are the distributions of $uu$-,$ud$- and $dd$-diquarks. 
$D_{u_v \rightarrow \pi}, D_{d_v \rightarrow \pi}$ and 
$D_{dd \rightarrow \pi}$ are the 
fragmentation functions of the valence $u$ and $d$ quarks and the 
$dd$ diquark into pions, respectively.  
The expressions of the quark and diquark
distributions and fragmentation functions employed in our calculations, 
obtained within the approach of ref.\ci{kaid2},  are given in 
the Appendix. At small $Q^2$ the cylindrical graph of fig.1b 
corresponds to one-Pomeron exchange in the $t$-channel, having 
the asymptotic behaviour $W^{{\widetilde \alpha}_P(0)-1}_X$. 
For the supercritical Pomeron the value of ${\widetilde \alpha}_P(0)$ 
is given 
by the relation $\Delta \equiv {\widetilde \alpha}_P(0)-1 \simeq 0.08$ 
\ci{kaid1,kaid2}. 
Comparison between the $W_X$-dependence of the cylindrical and
planar graphs of fig.1 shows that the contribution of the
planar graph can be neglected at large $W_X$ and not too large $Q^2$.   

The functions $\phi_1(x,Q^2)$ and $\phi_2(x,Q^2)$ describe the upper
vertices  of figs. 1a and 1b. In the case of neutrino-nucleon interaction
they are proportional to the charged electroweak current. In conclusion, 
the relativistic-invariant spectrum $z d^3\sigma/dx dz dp_t$ 
of pions produced in $\nu + N \rightarrow \mu^- + \pi^{\pm} + X$ processes 
can be written in the form
\be
z\ \frac{d^3\sigma}{dx dz dp_t} = F_P(x,Q^2;z,p_t)
\left( \frac{W_X}{s_0} \right)^{\alpha_R(0)-1} + F_C(x,Q^2;z,p_t)
\left( \frac{W_X}{s_o} \right)^{{\widetilde \alpha}_P(0)_1-1}\ ,
\label{20}
\ee
where $s_0 = 1.GeV^2$ is a parameter usually introduced in 
Regge theory in order to get the correct dimension of the matrix
elements or cross sections.
Eq.(\ref{20}) clearly shows that the cylindrical graph (fig. 1b) provides 
the main contribution to the spectrum at large $W_X$. 

The elementary spectrum $\rho_{ \nu N \rightarrow \mu^- \pi^{\pm} X}$
corresponding to backward production can be readily obtained from the above
$z d^3\sigma/dxdzdp_t$. Substituting into eq.({\ref{6}) one can 
finally evaluate the semi-inclusive spectrum for the case of
neutrino-nucleus scattering and compare to the available $\nu$-Ne 
data, taken at CERN by the WA 59 collaboration \ci{wa59}.

In order to compare to the data, the calculated spectrum has to be 
integrated over $x$ and $Q^2$. Since the main contribution to the integral
comes from the region of small $Q^2$ and large $\nu/s$ ($\nu$ 
denotes the neutrino energy loss, while $s = (q + k)^2$), implying in
turn large $W_X$, we have included in the calculation only the contribution 
of the cylindrical graph of fig. 1b. The calculated
$(E/\sigma)\rho_{\nu + Ne \rightarrow \mu^- + \pi + X}(p^2_\pi)$, 
where $p_\pi$ and $\sigma$ are the magnitude of the pion momentum and
the total cross section, respectively, is shown in fig. 2 together 
with the experimental 
data of ref.\ci{wa59}. For comparison, we also show (dashed line) 
the results of a calculation carried out using the spectral function
of ref.\ci{bf1}, which vanishes for $k >$ 0.8 GeV/c, to evaluate the 
distribution function of eq.(\ref{7}).
The large difference between the dashed and solid lines of fig. 2, particularly 
at high pion momenta, $p_\pi > 0.4$ GeV/c, indicates that 
the contribution of the kinematical region corresponding to 
nucleon momenta larger than $\sim$ 0.8 GeV/c dominates. The shape 
of the pion spectra at large 
momenta turns out to be determined by the 
asymptotic behaviour of the nucleon distribuition $f_A$ as a
function of the relativistic invariant variable 
$y = (M_A/m)(kq)/(P_A q)$. As a consequence, it appears that 
semi-inclusive lepton-nucleus and hadron-nucleus 
reactions in the kinematical region forbidden to free $\ell N$ or 
$h N$ scattering can be related to each other. 
    
In this paper we have applied the approach developed in ref.\ci{bfl1} 
to construct the distribution function  $f_A(y,Q^2)$ (see eq.(\ref{7})) 
in the region of very large $y$ ($y > 1.7$), dominated by short range
nuclear dynamics. Short range nucleon-nucleon correlations
are described in terms of overlapping distributions of three-quark colorless
objects. The conventional treatment of nucleon-nucleon correlations, 
based on the spectral function obtained 
within nonrelativistic many-body theory, cannot be employed to obtain 
$f_A(y,Q^2)$ at $y > 1.7$, the nucleon momenta involved being too large
$( > .8$ GeV/c). Our approach, originally proposed to describe 
hadron production in hadron-nucleus  processes refs.\ci{ekl1,ekl2,lyk}, 
quantitatively accounts for the spectra of backward pions produced 
in the $\nu + Ne  \rightarrow \mu^- + \pi^{\pm} + X$ process. Thus, our 
results suggest 
that the same mechanism is responsible for both hadron- and lepton-induced
emisssion of fast backward hadrons from nuclear targets.
         
\acknowledgments

This work has been encouraged and supported by the Russian Foundation
of Fundamental Research. We gratefully acknowledge many helpful 
discussions with D. Amati.

\appendix
\section{}

The distributions of valence quarks ($q_v(z)$) and
diquarks ($f_{qq}(z)$) ($q$ denotes either the $u$ or $d$ quark), exhibiting  
true Regge asymptotic at small $Q^2$, can be written according to \ci{kaid2}:
\be
q_v(z)=C_qz^{-\alpha_R(0)}(1 - z)^{\alpha_R(0)-2{\widetilde \alpha}_B(0)}\ ,
\label{A1}
\ee
and
\be
f_{qq}(z)=C_{qq}z^{\alpha_R(0)-2{\widetilde \alpha}_B(0)}(1-z)^{-\alpha_R(0)}\ .
\label{A3}
\ee
The coefficients $C_q$ and $C_{qq}$ are determined by the normalization 
conditions
\be
\int_0^1 q_v(z)dz =  \int_0^1 f_{qq}(z)dz = 1\ .
\label{A6}
\ee
The fragmentation functions have the form \ci{kaid2}
\be
D_{u_v \rightarrow \pi^+} = \frac{a_0}{z}(1-z)^{\alpha_R(0)+\lambda}\ ,
\label{A7}
\ee
\be
D_{u_v \rightarrow \pi^-}(z) = (1-z)D_{u_v \rightarrow \pi^+}(z)\ ,
\label{A8}
\ee
\be
D_{d_v \rightarrow \pi^+}(z) = D_{u_v \rightarrow \pi^-}(z)\ ,
\ee
\be
D_{d_v \rightarrow \pi^-}(z) = D_{u_v \rightarrow \pi^+}(z)\ ,
\label{A10}
\ee
\be
D_{uu \rightarrow \pi^+} = 
\frac{a_0}{z}(1-z)^{\alpha_R(0)-2{\widetilde \alpha}_B(0) + \lambda}\ ,
\label{A11}
\ee
\be
D_{uu \rightarrow \pi^-}(z) = (1-z)D_{uu \rightarrow \pi^+}(z)\ ,
\label{A12}
\ee
\be
D_{ud \rightarrow \pi^+}(z) = D_{ud \rightarrow \pi^-}(z) =
\frac{a_0}{z}[1+(1-z)^2](1-z)^{\alpha_R(0)-2{\widetilde \alpha}_B(0)
+\lambda}\ ,
\label{A13}
\ee
\be
D_{dd \rightarrow \pi^-}(z) = D_{uu \rightarrow \pi^+}(z)\ ,
\ee
and
\be
D_{dd \rightarrow \pi^+}(z) = D_{uu \rightarrow \pi^-}(z)i\ .
\label{a14}
\ee
In the above equations $\alpha_R(0)=0.5$ is the intercept of the Reggeon 
trajectory, ${\widetilde \alpha}_B(0)=-0.5$ is the intercept of the average 
baryonic trajectory, $a_0$ = 0.65 and $\lambda = 2\alpha'_R(0) <p_t^2> 
\simeq 0.5$, $\alpha'_R(0)$ and $<p_t^2>$ being the slope of the Reggeon 
trajectory and the average value of the transverse hadron momentum squared.

\begin{figure}
\caption{
Planar (a) and cylindrical (b) graphs contributing to the 
reaction $\nu + N \rightarrow \mu^- + h + X$. Diagrams (a) 
and (b) describe processes in which the incoming neutrino 
interacts with a valence quark or a sea quark (or antiquark), 
respectively. 
}

\label{fig1}
\end{figure}

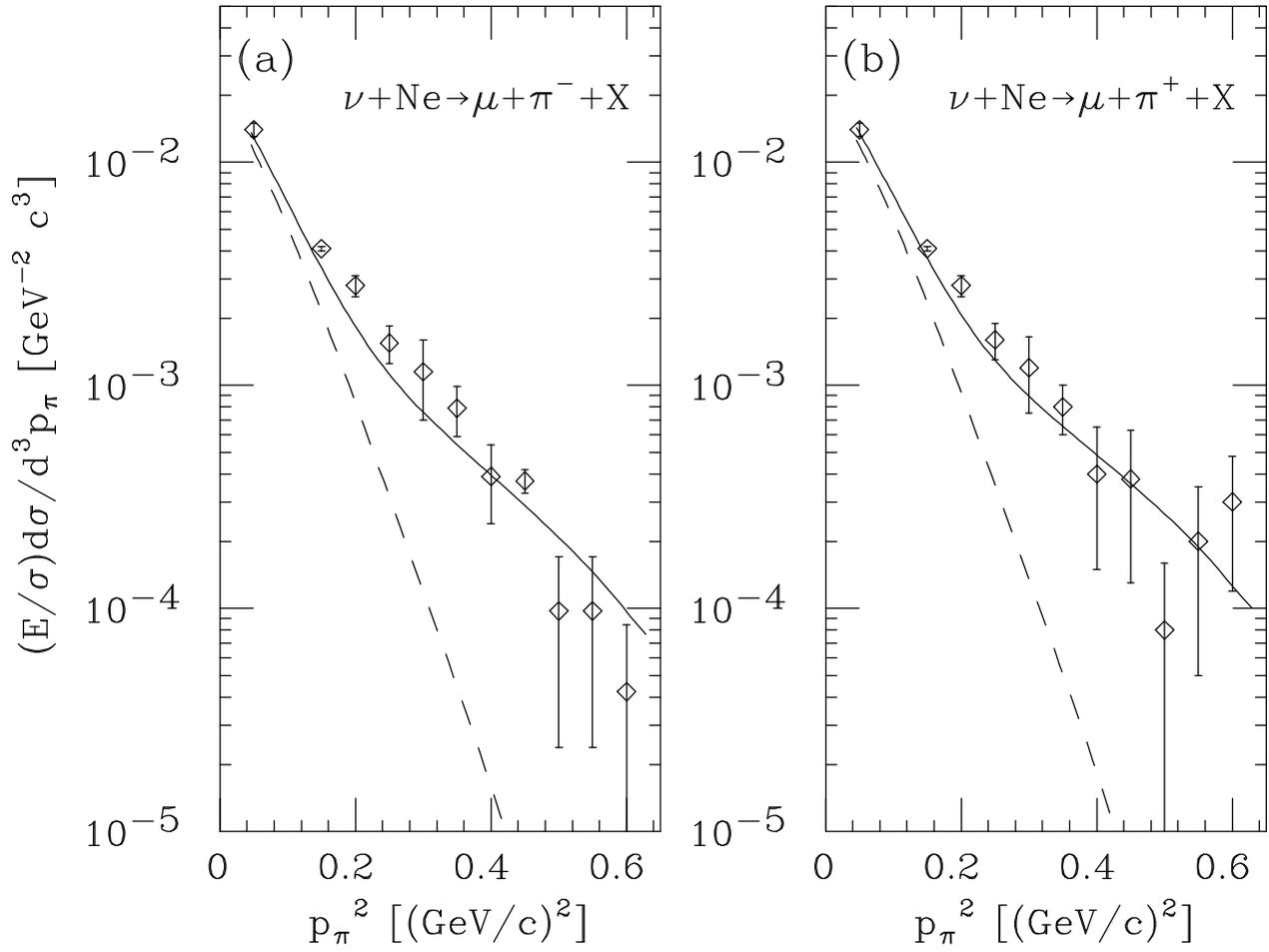
\begin{figure}
\caption{
(a): $p^2_\pi$-dependence of the spectrum 
of $\pi^-$-mesons
produced backward in the $\nu + Ne \rightarrow \mu^- + \pi^- X$ 
process. The solid
line has been obtained using the approach described in this paper, 
whereas the dashed curve shows the results of a calculation 
carried out using the nuclear spectral function $P(k)$ 
of ref.\protect\cite{bf1}. The experimental data are taken from 
ref.\protect\cite{wa59}. (b): same as in (a) but for $\pi^+$. 
}

\label{fig2}
\end{figure}

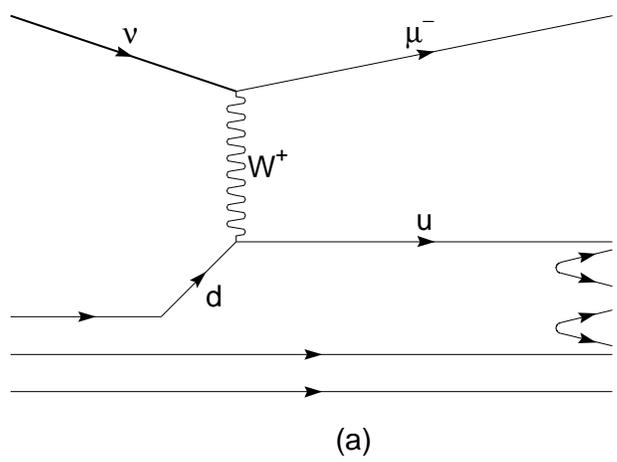
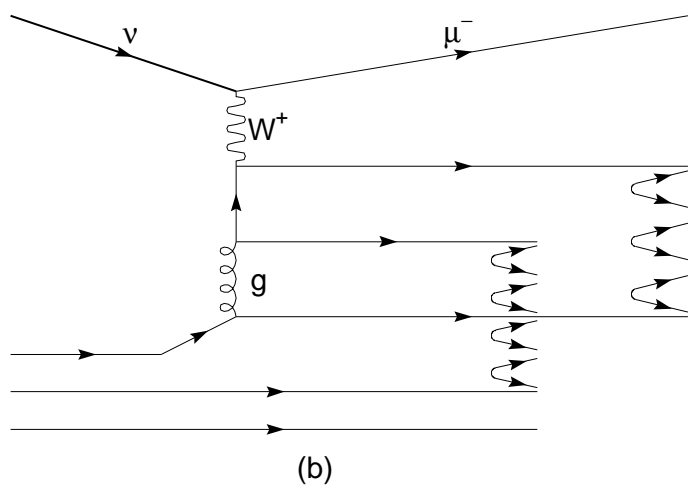
\end{document}